\documentclass[prb,superscriptaddress,amsmath,amssymb,floatfix,twocolumn,amsfonts]{revtex4-2}

\usepackage{times}
\usepackage[varg]{txfonts}
\usepackage{textcomp}
\usepackage{graphicx}
\usepackage{subfigure}
\usepackage{subeqnarray}
\usepackage{mathtools}
\usepackage{mathrsfs}
\usepackage{color}
\usepackage[colorlinks=true,citecolor=blue,urlcolor=blue,linkcolor=blue,hyperindex]{hyperref}
\usepackage{braket}
\usepackage{overpic}
\usepackage{ulem}
\usepackage[version=3]{mhchem}

\begin{document}

\title{Quasiparticles and optical conductivity in the mixed state of Weyl superconductors with unconventional pairing}

\author{Zhihai Liu}
\affiliation{School of Physics and Information Engineering, Guangdong University of Education, Guangzhou 510303, China}
\affiliation{College of Physics and Optoelectronic Engineering, Shenzhen University, Shenzhen 518060, China}

\author{Luyang Wang}
\email{wangly@szu.edu.cn}
\affiliation{College of Physics and Optoelectronic Engineering, Shenzhen University, Shenzhen 518060, China}


\begin{abstract}
Previous investigations have revealed that the Weyl superconductor (WeylSC), realized in a superconductor-topological insulator heterostructure, can exhibit the Landau levels (LLs) of Bogoliubov quasiparticles in the presence of a vortex lattice. Here, we investigate the low-energy quasiparticle (QP) excitations in the mixed state of heterostructure WeylSCs with unconventional pairing. We find that the spin-singlet $d$-wave pairing induces flat Dirac-LLs of Bogoliubov QPs, whereas the excitation spectra for the spin-triplet chiral $p$-wave pairing show noticeable dispersion, except for the chiral symmetry-protected, dispersionless, zeroth Landau level (ZLL). Distinct QP excitations in the vortex lattice of WeylSCs result in different optical responses, which are manifested as characteristic magneto-optical conductivity curves. We also show that, compared to the topologically protected, charge-neutral, localized Majorana zero mode (MZM), the chiral symmetry-protected ZLL is non-charge-neutral and delocalized. Both of these zero modes may be observed in the mixed state of a heterostructure topological superconductor.
\end{abstract}
 
\maketitle

\section{Introduction}\label{sec:intro}
In a magnetic field, low-energy QPs in most superconductors cannot condense into LLs. This is because, for first-kind superconductors, the magnetic field is completely expelled from the sample as a result of the Meissner effect; for second-kind superconductors, the magnetic field partially penetrates the sample, resulting in the mixed state where the Abrikosov vortex lattice is formed. The spatially varying supercurrents in the vortex lattice strongly scatter the QPs, obscuring the Landau level (LL) quantization~\cite{Melnikov.JPCM1999}. For example, in the mixed state of a single-band $d$-wave superconductor that exhibits Bogoliubov-Dirac (BD) nodes in the superconducting gap, the low-energy QPs remain in Dirac cone states~\cite{Franz.PRL2000,Vafek.PRB63.2001}, and the main effect of the magnetic field is the renormalization of the Fermi velocity~\cite{Yasui.PRL1999,Marinelli.PRB2000,Kopnin.PRB2000,Morita.PRL2001,Knapp.PRB2001,Vishwanath.PRL2001,Vafek.PRL2006,Melikyan.PRB2007}.

The WeylSC is a superconductor characterized by Bogoliubov-Weyl (BW) gap nodes~\cite{Meng.PRB2012,Meng.PRB2017,Nakai.PRB2020}, which may give rise to a distinct scenario. For instance, by using a heterostructure engineered by alternately stacking $s$-wave superconductor and Weyl semimetal (WSM) layers, Pacholski \textit{et al.} have revealed Dirac-LLs of Bogoliubov QPs in the mixed state of the heterostructure WeylSC~\cite{Pacholski.PRL2018}. At a BW point, the Dirac-LLs are proportional to $\sqrt{nB}$, with $n$ being the index of the LLs and $B$ the external magnetic field, where the dispersionless $n=0$ LL, located at $E=0$, is protected by chiral symmetry. These novel Bogoliubov LLs are quite similar to those in normal WSMs~\cite{TchoumakovPRL2016,YuZhi-MingPRL2016,ZhihaiLiuPRB2024}. This is because, unlike the situation in the single-band $d$-wave superconductor, the supercurrent velocity in the heterostructure WeylSC model always couples to the BW fermions as a vector potential. In addition, discrete Bogoliubov LLs have been studied in the mixed state of 2D heterostructure superconductors~\cite{Pacholski.PRL2021,ZhihaiLiu.arxiv2025}.

In fact, superconductor heterostructures, including those with conventional~\cite{FuLiangPRL2008,TanakaPRL2009,SauJayDPRL2010,QiXiao-LiangPRB2010,Mei-XiaoScience2012,XuSu-YangNatureP2014,XuJin-PengPRL2014} and unconventional~\cite{LinderPRL2010,LucignanoPRB2012,WangEryinNP2013} pairing, have been extensively studied in the context of topological superconductivity. The signal of the zero-bias peak, potentially caused by MZMs, has been observed experimentally in the vortex inside the topological insulator-superconductor heterostructure~\cite{XuJin-PengPRL2015,SunHao-HuaPRL2016}. By studying the proximity-induced superconductivity on the surface of a topological insulator (TI) layer deposited on a bulk unconventional superconductor, Linder \textit{et al.} demonstrated that QP excitations (in zero magnetic field) in the heterostructure differ qualitatively between spin-singlet and spin-triplet pairing~\cite{LinderPRL2010}: the excitations are gapped for the former, whereas they remain ungapped for the latter. As a result, both subgap bound states and Andreev reflection  associated with spin-triplet pairing are strongly suppressed. However, QPs in the WeylSC, realized in heterostructures with unconventional pairing, have not been thoroughly investigated, especially those in the mixed state.

By drawing an analogy with the MB model of WeylSCs~\cite{Meng.PRB2012,Meng.PRB2017}, unconventional pairing-induced Weyl superconductivity may be realized by alternately stacking TI (or WSM) and unconventional superconductor layers. In this work, we investigate low-energy QP excitations in the mixed state of heterostructure WeySCs with unconventional pairing and find that the spin-singlet $d$-wave pairing induces an ideal Dirac-LL structure, which is similar to that of the $s$-wave pairing~\cite{Pacholski.PRL2018,ZhihaiLiuPRB2024}. In contrast, QP bands for the spin-triplet chiral $p$-wave pairing exhibit strong dispersion rather than Dirac-LLs, except for the chiral symmetry-protected dispersionless ZLL. In addition, we reveal the distinctions between the MZM and the ZLL. Both of these zero-energy modes may emerge in the mixed state of heterostructure topological superconductors: (1) the MZM is charge-neutral, while the ZLL has a nonzero effective charge; (2) the MZM is localized at the vortex core, while the ZLL is delocalized; (3) the MZM is topologically protected, while the ZLL is protected by chiral symmetry.

Weyl superconductivity has also been proposed in stoichiometric materials such as \ce{URu_2Si_2}~\cite{KasaharaPRL2007,Goswami.arxiv2013}, \ce{UPt_3}~\cite{TouHPRL1998,GoswamiPRB2015}, \ce{UCoGe}~\cite{HuyNTPRL2007}, \ce{UTe_2}~\cite{JiaoNature2020,YuPRB2022}, and \ce{SrPtAs}~\cite{BiswasPKPRB2013,FischerPRB2014}. In these materials, the Weyl superconducting phase is typically realized through a chiral pairing state that breaks time-reversal symmetry. If the superconducting pairing occurs primarily between intra-band electrons and inter-band coupling can be ignored, intrinsic Weyl superconductivity can be described by a single-band superconductor model. Research on the excitation spectrum in the mixed state of a single-band WeylSC reveals the simultaneous presence of topologically protected MZMs and phase gradient-induced pseudo-LLs~\cite{ZhihaiLiu.arxiv2025}. In this work, we also investigate the magneto-optical conductivity of various Weyl superconducting systems, including the single-band WeylSC and heterostructure WeylSCs with unconventional pairing. These systems exhibit significantly different optical characteristics due to the distinct QP excitations.
\section{Model}\label{sec:Model}
\subsection{Intrinsic WeylSCs}\label{subsec:SpWeyl}
Considering the simplest scenario, i.e., that stoichiometric WeylSCs feature only intra-band superconducting pairing and that inter-band coupling can be ignored. Therefore, as mentioned above, intrinsic Weyl superconductivity can be described by a single-band superconductor model
\begin{eqnarray}
\mathscr{H}({\bf k})=
\begin{pmatrix}
\mathscr{H}_0({\bf k}) & \Delta({\bf k}) \\
\Delta^*({\bf k})  &  -\mathscr{H}^*_0(-{\bf k}) 
\end{pmatrix},
\label{Ham_SPWeyl}
\end{eqnarray}
where the single-particle lattice Hamiltonian is given by
\begin{eqnarray}
\mathscr{H}_0({\bf k}) &= \cos{k_za_0} + \cos{k_xa_0} + \cos{k_ya_0} - \mu,
\end{eqnarray}
with $\mu$ as the chemical potential and $a_0$ as the lattice constant (setting $a_0 \equiv 1$). The Weyl superconductivity described in Eq.~(\ref{Ham_SPWeyl}) can be realized through chiral pairing, such as the proposed chiral $p$-wave pairing in \ce{UTe_2}~\cite{YuPRB2022}; the chiral $d$-wave pairing in \ce{URu_2Si_2}~\cite{Goswami.arxiv2013}; and the chiral $f$-wave pairing in \ce{UPt_3}~\cite{GoswamiPRB2015}. The chiral $d$- and $f$-wave pairing states generate concomitant nodal lines or double-Weyl nodes. Here, we focus only on the low-energy physics near the BW nodes, considering the simple chiral $p$-wave pairing state $\Delta({\bf k})=\Delta (\sin{k_x}-i\sin{k_y})$, where $\Delta=\Delta_0e^{i\phi({\bf r})}$ with $\phi({\bf r})$ being the globally coherent superconducting phase and $\Delta_0$ being the pairing magnitude. 

\begin{figure}[ht]
  \centering
  \subfigure{\includegraphics[width=2.5in]{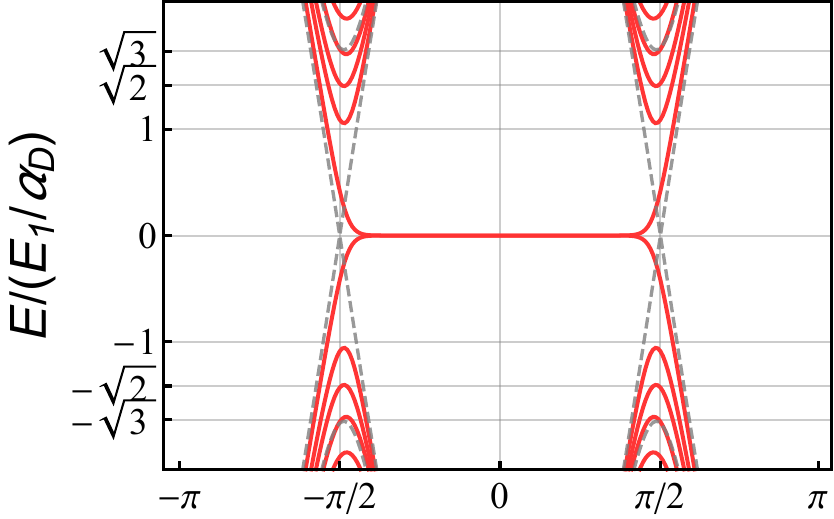}}
  \caption{The excitation spectrum of the intrinsic WeylSC along the momentum $k_z$ ($k_x=k_y=0$) in zero magnetic field (dashed gray lines) and in the vortex lattice (solid red lines), with the magnetic length $l_B=22$, $\Delta_0=1$ and $\mu=2$.} \label{fig:SpLLz}
\end{figure}
To clearly understand the magneto-optical conductivity of intrinsic WeylSCs, we provide a brief overview of the QP excitations in the mixed state of the single-band WeylSC. These low-energy QPs are described by a continuum Hamiltonian ($\hbar, c \equiv 1$, electron charge $e$, mass $m$)
\begin{eqnarray}
\mathcal{H}_{WI} = v_\Delta \sum_{\alpha=x,y} \chi_\alpha \sigma_\alpha  + \tfrac{1}{2m} (k_z^2-\mathcal{K}^2) \sigma_z + \mathcal{H}_1,
\label{TiltedWeyl}
\end{eqnarray}
where $\boldsymbol{\sigma}$ is the vector of Pauli matrices, $\boldsymbol{\chi} = {\bf k} + {\bf a}$, $v_F$ and $v_\Delta=\Delta_0/k_F$ are the Fermi velocity and gap velocity, respectively, while $k_F$ represents the Fermi momentum. Here, we retain the quadratic term $\mathcal{H}_1$ to better describe the MZM physics, with
\begin{eqnarray}
\mathcal{H}_1=
\begin{pmatrix}
\tfrac{1}{2m}(\chi_{||} + m\boldsymbol{v}_s)^2 - \mu   & 0 \\
0  &  -\tfrac{1}{2m}(\chi_{||} - m\boldsymbol{v}_s)^2 + \mu
\end{pmatrix},
\end{eqnarray}
the subscript $||$ indicates the $k_x-k_y$ plane. The gauge field ${\bf a}= \tfrac{1}{2}\nabla \phi $ and the supercurrent velocity $\boldsymbol{v}_s =(\tfrac{1}{2}\nabla \phi - e{\bf A})/m$, where ${\bf A}$ is the external magnetic vector potential that generates a uniform magnetic field ${\bf B}=B_0\hat{z}$.

At the BW points ($k_z=\pm \mathcal{K}$), the quadratic term $\mathcal{H}_1$ can be ignored, and Hamiltonian~(\ref{TiltedWeyl}) respects chiral symmetry, while the supercurrent velocity (where the external magnetic vector potential enters) does not couple to the BW fermions. However, the phase gradient ${\bf a}$ couples to the BW fermions as an effective vector potential, thereby inducing pseudo-LLs $E_n=\sqrt{n} E_1/\alpha_D$ near the BW nodes, where $E_1=2\sqrt{\pi} v_F/l_B$ and $\alpha_D=v_F/v_\Delta$. On the other hand, when deviating from the BW points, the quadratic term breaks the chiral symmetry of Hamiltonian~(\ref{TiltedWeyl}). The single-band WeylSC model~(\ref{Ham_SPWeyl}) is essentially a 2D spinless $p_x-ip_y$ superconductor with a parameter $k_z$. Therefore, topologically protected MZMs emerge as long as the system is in the weak-pairing phase~\cite{VolovikJETPL1999,Read.PRB2000}, as shown in Fig.~\ref{fig:SpLLz}. 
\subsection{Heterostructure WeylSCs with unconventional pairing}\label{subsec:TdpWeyl}
Taking the TI-superconductor heterostructure as an example, the proximity effect may induce either conventional or unconventional superconducting pairing on the surface of the TI layers, depending on the pairing symmetry of the superconductor layers. Proximity-induced superconductivity can be described by a Bogoliubov–de Gennes (BdG) Hamiltonian
\begin{eqnarray}
\mathscr{H}({\bf k})=
\begin{pmatrix}
\mathscr{H}_0({\bf k}) & \underline{\Delta}({\bf k}) \\
\underline{\Delta}^*({\bf k})  &  -\mathscr{H}^*_0(-{\bf k}) 
\end{pmatrix},
\label{Ham_TBWeyl}
\end{eqnarray}
where $\underline{\Delta}({\bf k})$ represents a $2\times 2$ gap matrix. The normal-state Hamiltonian reads
\begin{eqnarray}
\begin{aligned}
\mathscr{H}_0({\bf k}) =&  \sum_{\alpha=x,y} \sigma_\alpha\sin{k_\alpha} + \sigma_z\cos{k_z} - \mu\sigma_0 \\
&+ \sigma_z M_0\sum_{\alpha=x,y}(1-\cos{k_\alpha}).
\end{aligned}
\label{Ham_NS}
\end{eqnarray}
Here, we also include a quadratic $M_0$ term (setting $M_0=1$ hereafter) to eliminate fermion doubling at specific points, such as ($\pi, \pi$). The gap matrix $\underline{\Delta}({\bf k})$ depends on both the orbital and spin symmetry of the Cooper pair. For spin-singlet pairing, such as $s$-wave and $d$-wave, one has $\underline{\Delta}({\bf k})= \Delta({\bf k})i\sigma_y$ with $\Delta({\bf k}) = \Delta$ for $s$-wave and $\Delta({\bf k}) = \Delta(\cos{k_x}-\cos{k_y})$ for $d_{x^2-y^2}$-wave. For spin-triplet pairing, the gap matrix reads $\underline{\Delta}({\bf k})= ({\bf d_k}\cdot \boldsymbol{\sigma})i\sigma_y$~\cite{KallinPPP2016,GhoshJPCM2021}. Here, we consider a simple chiral $p$-wave pairing state with ${\bf d_k} = \Delta({\bf k}) \hat{\bf z}$ and $\Delta({\bf k}) = \Delta (\sin{k_x} - i\sin{k_y})$. Thus, we have $\underline{\Delta}({\bf k})= \Delta (\sin{k_x} - i\sin{k_y})\sigma_x$, which describes an opposite-spin triplet pair.

The QP excitation and magneto-optical conductivity of heterostructure WeylSCs with $s$-wave pairing have been investigated~\cite{Pacholski.PRL2018,ZhihaiLiuPRB2024}. Therefore, in this work, we primarily focus on the unconventional $d$-wave and chiral $p$-wave pairing. The low-energy physics near a BW node in Eq.~(\ref{Ham_TBWeyl}) can be more clearly understood using a continuum BdG Hamiltonian, in which the normal-state Hamiltonian is given by
\begin{eqnarray}
H_0({\bf k}) =  v_F\sum_{\alpha=x,y,z} \sigma_\alpha k_\alpha - \mu\sigma_0.
\label{LHam_NS}
\end{eqnarray}
The gap matrix $\underline{\Delta}({\bf k})= \tfrac{\Delta}{k_F^2} (k_x^2-k_y^2)i\sigma_y$ for the $d_{x^2-y^2}$ pairing and $\underline{\Delta}({\bf k})= \tfrac{\Delta}{k_F} (k_x - ik_y)\sigma_x$ for the spin-triplet chiral $p$-wave pairing.

\begin{figure}[ht]
  \centering
  \subfigure{\includegraphics[width=1.6in]{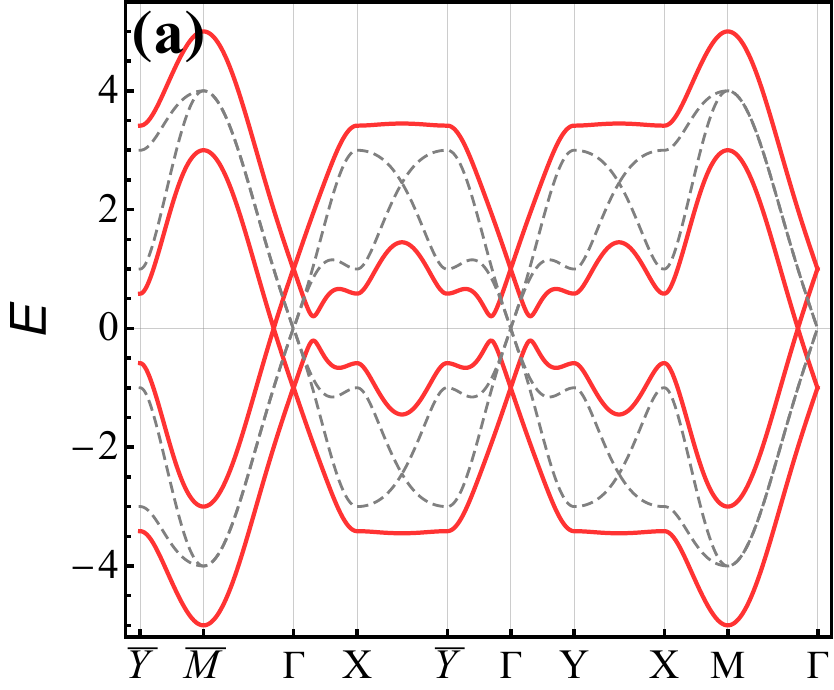}\label{fig:bandd}}~~
  \subfigure{\includegraphics[width=1.6in]{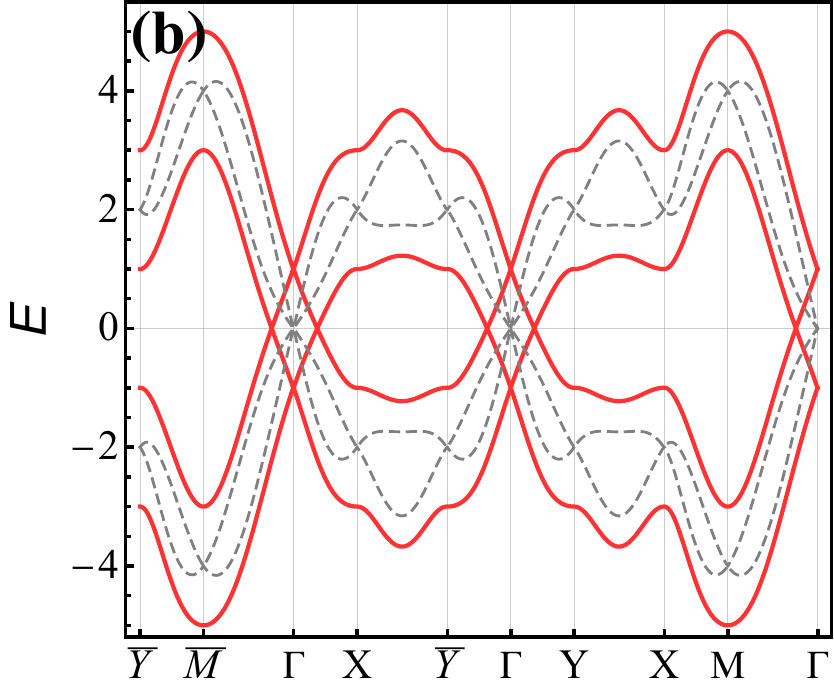}\label{fig:bandp}}
  \caption{Excitation spectra of WeylSC model~(\ref{Ham_TBWeyl}) in zero magnetic field, with $k_z=\tfrac{\pi}{2}$ and $\Delta_0=0.5$. (a) $d_{x^2-y^2}$ pairing. (b) $p_x-ip_y$ pairing. The solid red and dashed gray lines denote $\mu=1$ and $\mu=0$, respectively.} \label{fig:BandS}
\end{figure}

Considering the $k_z=0$ subspace of Eq.~(\ref{Ham_TBWeyl}), the continuum Hamiltonian for spin-singlet $d_{x^2-y^2}$-wave pairing yields eigenvalues 
\begin{eqnarray}
\epsilon_{{\bf k}_{||}} = \pm \sqrt{(v_F|{{\bf k}_{||}}| \pm \mu)^2 + |\Delta({\bf k})|^2}.
\label{Eig_singlet}
\end{eqnarray}
The excitation spectrum Eq.~(\ref{Eig_singlet}) exhibits two degenerate BD nodes located at ${\bf k}_{||}=0$ when $\mu=0$, while there are four BD nodes on the Fermi surface, located at ($\pm\mu/\sqrt{2}v_F$, $\pm\mu/\sqrt{2}v_F$), due to the nodal lines $\Delta(k_x=\pm k_y) =0$ of the $d_{x^2-y^2}$ pairing when $\mu \neq 0$, as shown in Fig.~\ref{fig:bandd}. Compared to the nodes at the point ${\bf k}_{||}=0$, the BD nodes on the Fermi surface do not induce chiral symmetry-protected ZLL in the presence of a vortex lattice, which is similar to the behavior of those in single-band $d$-wave superconductors, as revealed below.

For the spin-triplet chiral $p$-wave gap matrix $\underline{\Delta}({\bf k})= \tfrac{\Delta}{k_F} (k_x - ik_y)\sigma_x$, the eigenvalues in the $k_z=0$ subspace read
\begin{eqnarray}
\epsilon_{{\bf k}_{||}} = \pm v_F|{\bf k}_{||}| \pm \sqrt{\mu^2 + |\Delta({\bf k})|^2}.
\label{Eig_triplet}
\end{eqnarray}
Remarkably, the superconducting order parameter simply renormalizes the chemical potential, and a nodal ring emerges in the spectrum Eq.~(\ref{Eig_triplet}) when $\mu \neq 0$. When $\mu=0$, we have $\epsilon_{{\bf k}_{||}}|_{\mu=0} = \pm (v_F \pm v_\Delta)|{\bf k}_{||}|$; two BD cones with different Fermi velocities emerge at the point ${\bf k}_{||}=0$. Alternatively, we obtain an unconventional QP that is similar to the pseudospin-$\tfrac{3}{2}$ fermion in semimetals~\cite{HsiehPRB2014,Barry_Science2016,EzawaPRB2016}, as shown in Fig.~\ref{fig:bandp} by the dashed gray lines. In fact,  Linder \textit{et al.} have indicated that the anomalous ungapped dispersion Eq.~(\ref{Eig_triplet}) can be obtained for any triplet symmetry, and the results for singlet and triplet pairing differ qualitatively~\cite{LinderPRL2010}.

\section{Quasiparticles in the mixed state}\label{sec:CS&LLs}
In theory, WeylSCs are second-kind superconductors due to the BW gap nodes. Indeed, the vortex lattice has been observed experimentally in the TI-superconductor heterostructure~\cite{XuJin-PengPRL2014,XuJin-PengPRL2015,SunHao-HuaPRL2016}. Here, we consider low and intermediate magnetic fields, $H_{c1}< B_0\ll H_{c2}$, where $H_{c1}$ and $H_{c2}$ are the lower and upper critical fields, respectively. In this regime, vortex cores comprise a negligible fraction of the sample. Therefore, the magnitude $\Delta_0$ of the order parameter is approximately uniform throughout the sample, while the superconducting phase $\phi({\bf r})$ is strongly position-dependent.
\subsection{Chiral symmetry}\label{subsec:ChiralS}
In a magnetic field, the continuum Hamiltonian near a BW node in Eq.~(\ref{Ham_TBWeyl}) is written as
\begin{eqnarray}
H({\bf k})=
\begin{pmatrix}
H_0({\bf k}-e{\bf A}) & \underline{\Delta}({\bf k}) \\
\underline{\Delta}^*({\bf k}) & -H_0^*(-{\bf k}-e{\bf A})
\end{pmatrix},
\label{EQ:HamMix}
\end{eqnarray}
where $H_0$ is given in Eq.~(\ref{LHam_NS}). Here, the gap matrix $\underline{\Delta}({\bf k})$ is expressed in terms of operators, $\underline{\Delta}({\bf k}) = \hat{\Delta}_d i\sigma_y$ for $d_{x^2-y^2}$ pairing and $\underline{\Delta}({\bf k}) = \hat{\Delta}_p\sigma_x$ for $p_x-ip_y$ pairing, with $\hat{\Delta}_d = k_F^{-2} \{k_x, \{k_x, \Delta \} \}- k_F^{-2} \{k_y, \{k_y, \Delta \} \}$ and $\hat{\Delta}_p = k_F^{-1} \{k_x-ik_y, \Delta\}$, where ${\bf k}=-i\boldsymbol{\nabla}$, and the operators are symmetrized by $\{a,b\}=\tfrac{1}{2}(ab+ba)$. Following the method outlined in Ref.~\onlinecite{Pacholski.PRL2018}, the phase factors $e^{i\phi}$ in the gap matrix $\underline{\Delta}({\bf k})$ can be removed from the off-diagonal components and incorporated into the single-particle Hamiltonian $H_0$. This is accomplished by a gauge transformation (Anderson gauge) $H \to \mathcal{U}^\dagger H \mathcal{U}$ with~\cite{Anderson.arxiv1998}
\begin{eqnarray}
\mathcal{U} =
\begin{pmatrix}
e^{i\phi} & 0 \\
0 & 1
\end{pmatrix}.
\end{eqnarray}
 The transformed Hamiltonian reads
\begin{eqnarray}
\mathcal{H}=
\begin{pmatrix}
H_0(\boldsymbol{\chi}+m\boldsymbol{v}_s) & \underline{\Delta_0}(\boldsymbol{\chi}) \\
\underline{\Delta_0}^*(\boldsymbol{\chi}) & -H_0^*(-\boldsymbol{\chi}+m\boldsymbol{v}_s)
\end{pmatrix},
\label{EQ:HamGTrd}
\end{eqnarray}
where  $\underline{\Delta_0}(\boldsymbol{\chi})= \tfrac{\Delta_0}{k_F^2} (\chi_x^2-\chi_y^2)i\sigma_y$ for $d_{x^2-y^2}$ pairing and $\underline{\Delta_0}(\boldsymbol{\chi})= \tfrac{\Delta_0}{k_F} (\chi_x - i\chi_y)\sigma_x$ for $p_x-ip_y$ pairing.

For spin-singlet $d_{x^2-y^2}$-wave pairing, Hamiltonian~(\ref{EQ:HamGTrd}) can be block diagonalized using a unitary transformation $\mathcal{V}_S=\exp{(\tfrac{1}{2}i\vartheta \nu_y \sigma_x)}$~\cite{PBaireuther.NJP2017}, with $\tan \vartheta = -\tfrac{\Delta_0}{v_F k_z}(\chi_x^2-\chi_y^2)$, $\vartheta \in (0, \pi)$, and $\boldsymbol{\nu}$ is the vector of Pauli matrices that acts on the electron-hole index. The block-diagonalized Hamiltonian describes two BW cones that are degenerate in momentum but experience different effective magnetic vector potentials, we have (setting $\mu=0$)~\cite{HamiltSI}
\begin{eqnarray}
\mathcal{H}_{S\pm} = v_F \sum_{\alpha=x,y}(\chi_\alpha \pm mv_{s,\alpha}\cos{\vartheta})\sigma^\pm_\alpha - \Gamma_{k_z}\sigma_z,
\label{HamSDBW}
\end{eqnarray}
where $\Gamma_{k_z}=\sqrt{\Delta_0^2(\chi_x^2-\chi_y^2)^2 + v_F^2k_z^2}$, $\sum_{{\alpha=x,y}} \mathcal{R}_\alpha \sigma^\pm_\alpha = \mathcal{R}_x \sigma_x \pm \mathcal{R}_y \sigma_y$. Significantly, at the BW point, $\mathcal{H}_{S\pm}$ anticommute with $\sigma_z$, thereby respecting chiral symmetry and resulting in dispersionless ZLLs. The singlet pairing potential $\Delta$ only alters the effective magnetic vector potential (and the position of nodes for $s$-wave pairing). 

For spin-triplet $p_x-ip_y$ pairing, Hamiltonian~(\ref{EQ:HamGTrd}) can be block diagonalized through a unitary transformation $H \to \mathcal{V}_T^\dagger H \mathcal{V}_T$, with~\cite{LSCederbaumJPMG1989}
\begin{eqnarray}
\mathcal{V}_T = \tfrac{\sqrt{2}}{2}
\begin{pmatrix}
1 & 0 & e^{-2i\theta} & 0\\
0 & 1 & 0 & 1 \\
-e^{2i\theta} & 0 & 1 & 0 \\
0 & -1 & 0 & 1
\end{pmatrix},
\end{eqnarray}
where $\tan{\theta}=\chi_y/\chi_x$, $\theta \in (0, \pi)$. When $\mu=0$, the block-diagonalized matrix includes two decoupled $2\times 2$ Hamiltonians, given by~\cite{HamiltSI}
\begin{eqnarray}
\begin{aligned}
\mathcal{H}_{T\pm} = & \sum_{\alpha=x,y}[(v_F\mp v_\Delta) \chi_\alpha + mv_F\tilde{v}_{\pm,s,\alpha}]\sigma^\pm_\alpha \\
&+ v_Fk_z\sigma_z,
\end{aligned}
\label{HamCPBW}
\end{eqnarray}
where $\tilde{\boldsymbol{v}}_{\pm,s} = e^{-i\theta}(iv_{s,x}\sin{\theta}, \pm v_{s,y}\cos{\theta}, 0)$. Hamiltonians $\mathcal{H}_{T\pm}$ describe two BW cones that feature unequal Fermi velocities $v_F\pm v_\Delta$, respectively. Compared to the singlet pairing case, the triplet pairing potential renormalizes the Fermi velocities. Hamiltonian~(\ref{HamCPBW}) anticommute with $\sigma_z$ when $k_z=0$, and thus may host the chiral symmetry-protected ZLL.
\subsection{Excitation spectrum in the vortex lattice}\label{subsec:LLinLattice}
\begin{figure}[ht]
  \centering
  \subfigure{\includegraphics[width=1.62in]{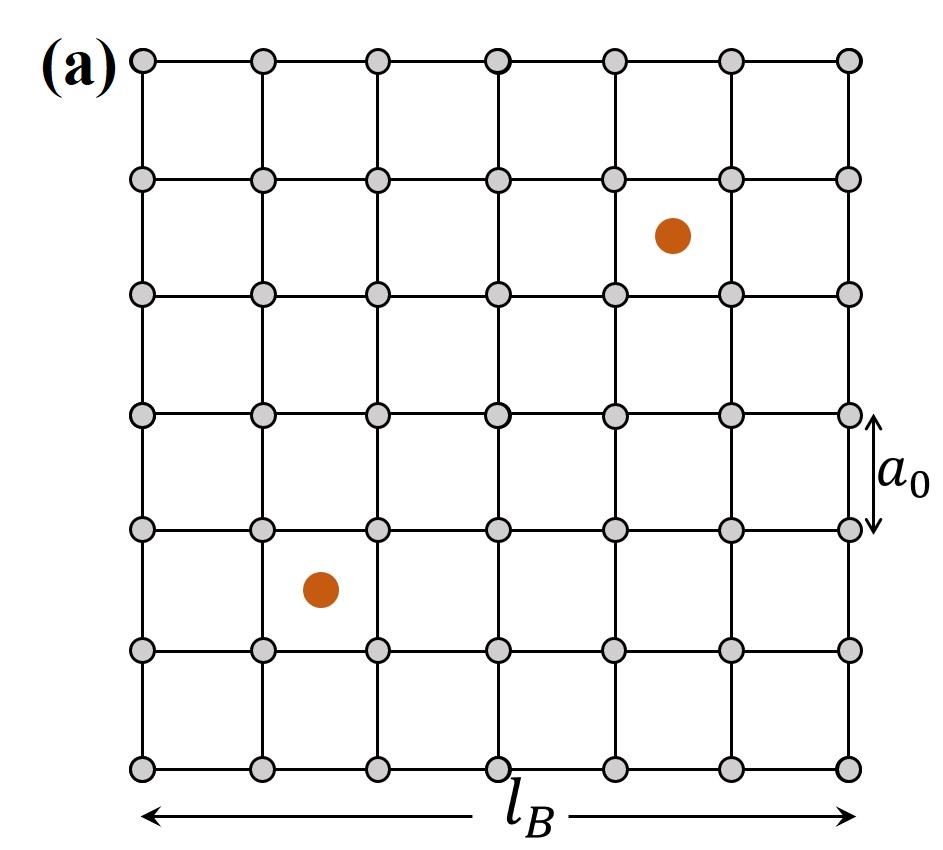}\label{fig:Lattice}}~~
  \subfigure{\includegraphics[width=1.58in]{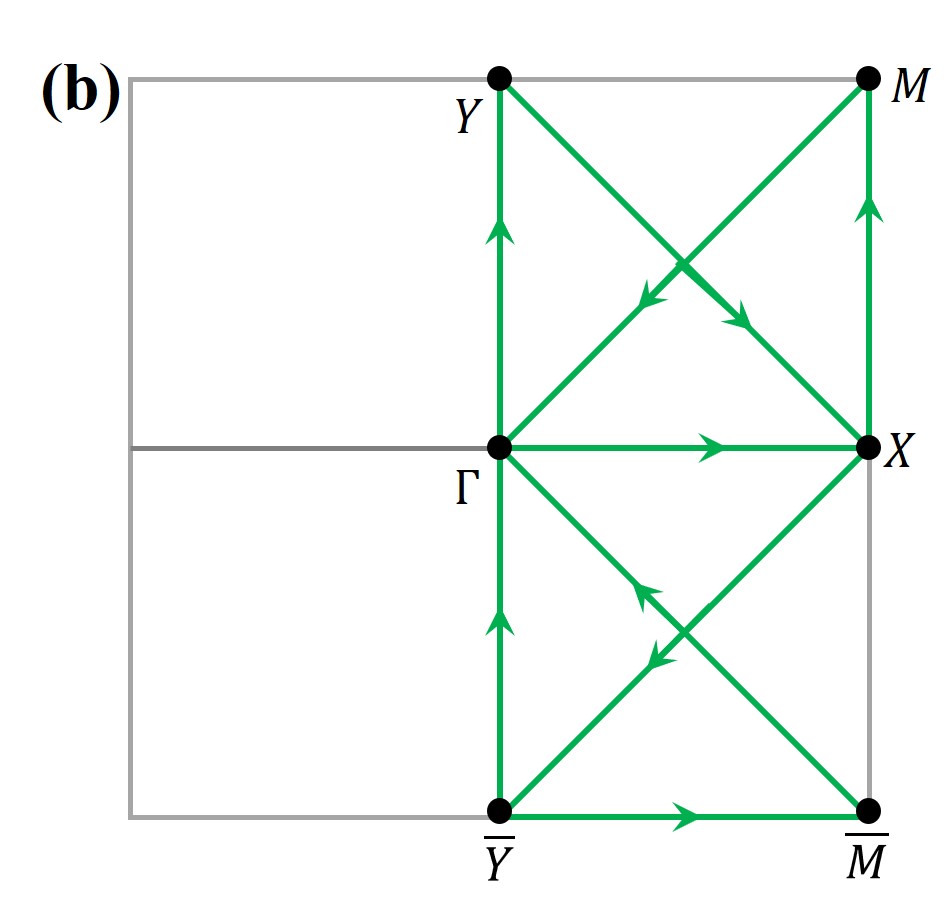}\label{fig:BZ}}
  \caption{(a) Magnetic unit cell $l_B \times l_B$ containing two vortices (orange solid circle) with $l_B=6a_0$. (b) A momentum path (green lines) in the first Brillouin zone of the vortex lattice.}
\end{figure}

Tight-binding calculations for the excitation spectra in the mixed state of WeylSCs are performed using a vortex lattice, as shown in Fig.~\ref{fig:Lattice}. In the $x-y$ plane, a square vortex lattice is formed with a magnetic unit cell $l_B\times l_B$, where the magnetic length is defined as $l_B=\sqrt{\phi_0/B_0}$, with the flux quantum $\phi_0=hc/e$. Here, we consider the simple case $l_B=(4N+2)a_0$ with $N$ as a positive integer. Each magnetic unit cell contains two evenly distributed quantized vortices represented by the orange solid circles, located at ($l_B/4$, $l_B/4$) and ($3l_B/4$, $3l_B/4$), respectively, each vortex carries a flux $hc/2e$. 

In the vortex lattice, while the vortex positions are periodic, the real-space lattice Hamiltonian $\hat{\mathcal{H}}({\bf r})$ is invariant only if the discrete translations are followed by a gauge transformation (magnetic translations). However, $\hat{\mathcal{H}}({\bf r})$ can be transformed into a periodic Hamiltonian by a singular gauge transformation $\mathcal{U}=\exp \{i\sigma_z \phi ({\bf r})/2\}$. It should be noted that the gauge transformation $\mathcal{U}$ is not single-valued due to the $2\pi$ winding of the superconducting phase around each vortex, nor is the transformed Hamiltonian. The  issue of multiple-valuedness can be handled by introducing compensating branch cuts~\cite{Vafek.PRB63.2001,Vafek.PRL2006,WangLuyang.PRB2013,ZhihaiLiuPRB2024,ZhihaiLiu.arxiv2025}. The estimation of the superconducting phase $ \phi ({\bf r})$ on a vortex lattice is discussed in Ref.~\onlinecite{Melikyan.PRB2007}, where it is expressed in closed form using the Weierstrass sigma function. Alternatively, it can also be calculated via a set of equations, as demonstrated in Ref.~\onlinecite{Pacholski.PRL2018}. 

\begin{figure}[ht]
  \centering
  \subfigure{\includegraphics[width=1.6in]{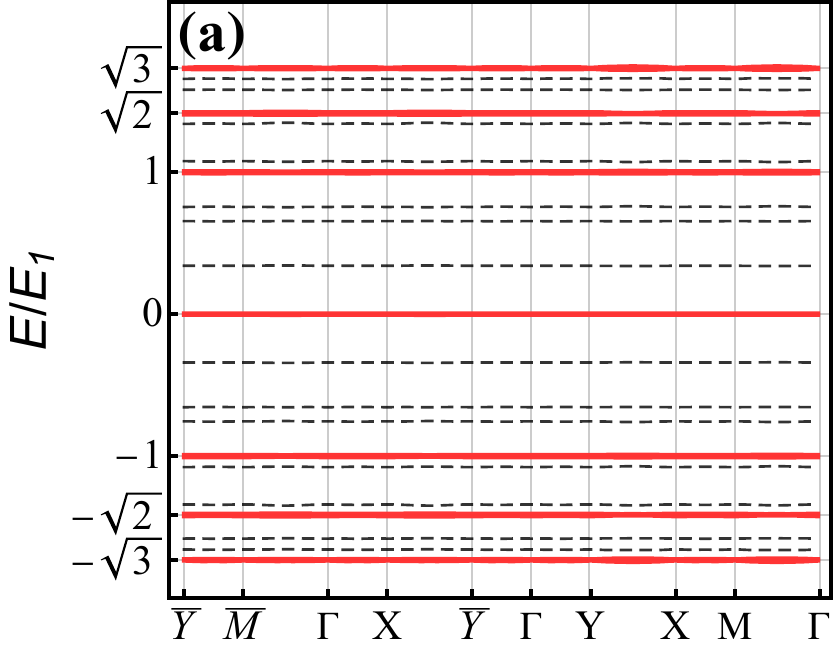}\label{fig:LLd}}~~
  \subfigure{\includegraphics[width=1.6in]{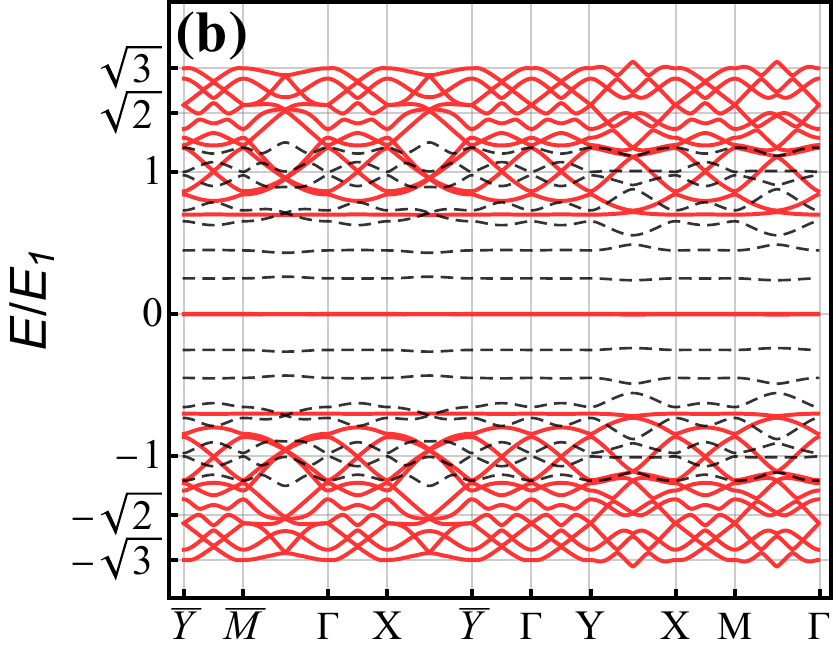}\label{fig:LLp}}
  \caption{Excitation spectra in the vortex lattice of heterostructure WeylSCs, with $l_B=22$, $k_z=\tfrac{\pi}{2}$ and $\Delta_0=0.5$. (a) $d_{x^2-y^2}$ pairing. (b) $p_x-ip_y$ pairing. The solid red and dashed gray lines indicate $\mu=0$ and $\mu=0.1$, respectively.}\label{fig:LLs}
\end{figure}

The numerically calculated excitation spectra in the $k_z=\tfrac{\pi}{2}$ subspace are shown in Fig.~\ref{fig:LLs}. When the chemical potential $\mu=0$, the low-energy Hamiltonian at BW points for both the $d_{x^2-y^2}$ and $p_x-ip_y$ pairing states respects chiral symmetry, as elucidated by Eqs.~(\ref{HamSDBW}) and~(\ref{HamCPBW}). Consequently, dispersionless ZLLs emerge at zero energy, as shown in Fig.~\ref{fig:LLd} and~\ref{fig:LLp}, respectively. The excitation spectrum in the vortex lattice for the $d_{x^2-y^2}$ pairing exhibits ideal Dirac-LLs $E_n=\sqrt{n}E_1$. In contrast, in the spectrum of the $p_x-ip_y$ pairing state, only the $n=0$ level is dispersionless, which is protected by chiral symmetry, while the other levels ($n>0$) exhibit noticeable dispersion. Indeed, the LLs with $n>0$ are not symmetry-protected.

When $\mu \neq 0$, as elucidated by Eqs.~(\ref{Eig_singlet}) and~(\ref{Eig_triplet}) (also see Fig.~\ref{fig:BandS}), the $d_{x^2-y^2}$ pairing generates four BD gap nodes, while the $p_x-ip_y$ pairing state exhibits a nodal ring. However, the nonzero chemical potential causes ZLLs in both pairing states to deviate from zero energy, as shown by the dashed gray lines in Fig.~\ref{fig:LLs}, which is similar to what occurs in Dirac semimetals with nonzero chemical potential. Significantly, the BD nodes on the Fermi surface in the $d_{x^2-y^2}$ pairing state, which are located at zero energy, do not result in Landau quantization due to the absence of ZLLs with $E=0$, as indicated by the dashed gray lines in Fig.~\ref{fig:LLd}. This demonstrates that the derivation of nodes in superconductors is a key ingredient for Landau quantization of Bogoliubov QPs. 
\section{Chiral symmetry-protected ZLL and topologically protected MZM}\label{subsec:MZM&ZLL}
In the $k_z=0$ subspace of Hamiltonian~(\ref{EQ:HamMix}), if we consider an isotropic $s$-wave pairing $\Delta({\bf k}) = \Delta_0e^{i\phi}$ and apply an in-plane supercurrent with Cooper pair momentum $\mathit{\boldsymbol{K}}$ in the $x$ direction, we obtain the 2D topological superconductor model proposed by Pacholski \textit{et al.} (abbreviated as the PB model hereafter)~\cite{Pacholski.PRL2021}. In the magnetic vector potential ${\bf A}$, the Hamiltonian of the PB model is written as 
\begin{eqnarray}
H ({\bf k}_{||}) =
\begin{pmatrix}
\mathit{K}\sigma_x + H_0({\bf k}_{||} - e{\bf A})   & \Delta_0e^{i\phi} i\sigma_y \\
- \Delta_0e^{-i\phi} i\sigma_y & -\mathit{K}\sigma_x - H^*_0(-{\bf k}_{||} - e{\bf A})
\end{pmatrix}.
\label{EQ:HamPB}
\end{eqnarray}
When $\mu=0$, Hamiltonian~(\ref{EQ:HamPB}) satisfies $\Lambda^\dagger H \Lambda = -H$ with $\Lambda = diag (1,-1,1,-1)$, thereby respecting chiral symmetry. In zero magnetic field, the excitation spectrum of the PB model exhibits a full gap when Cooper pair momentum $\mathit{K}<\Delta_0$ (confined phase) and two BD nodes located at $(\pm \sqrt{\mathit{K}^2-\Delta_0^2}, 0)$ when $\mathit{K}>\Delta_0$ (deconfined phase). The two BD nodes are at $E=0$ when $\mu=0$ and move up and down, respectively, when $\mu\neq 0$. 

The excitation spectrum in the mixed state of the PB model is calculated using a vortex lattice configuration. The fully gapped confined phase exhibits zero-energy modes (which disappear when $M_0=0$ in Eq.~(\ref{Ham_NS})), which are topologically protected rather than protected by chiral symmetry. This is because these zero-energy modes remain at $E=0$ even if $\mu \neq 0$, where the chiral symmetry of Hamiltonian~(\ref{EQ:HamPB}) is broken. These zero-energy modes in the confined phase are precisely the topologically protected MZMs. In contrast, the BD nodes in the deconfined phase induce Dirac-LLs given by $E_n=\sqrt{n}E_1$, including the dispersionless ZLL. However, the ZLL deviates from zero energy when $\mu \neq 0$ due to the breaking of chiral symmetry (see Fig. 3 in Ref.~\onlinecite{Pacholski.PRL2021}), which is similar to the scenario in heterostructure WeylSCs, as shown in Fig.~\ref{fig:LLs}.

\begin{figure}[ht]
  \centering
  \subfigure{\includegraphics[width=1.6in]{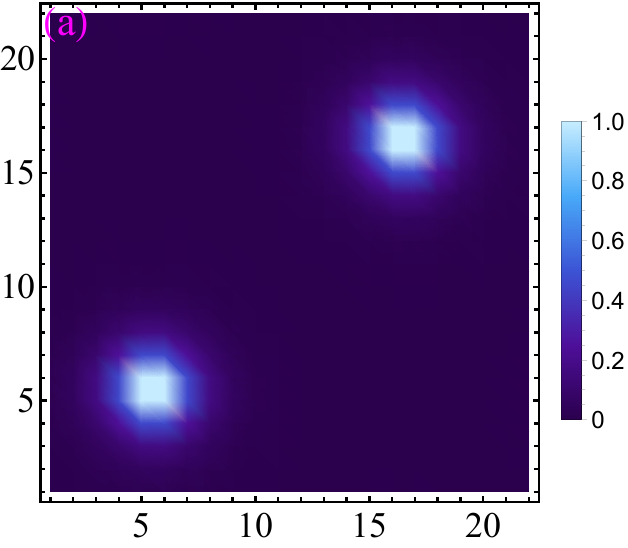}\label{fig:IntsK0}}~~
  \subfigure{\includegraphics[width=1.6in]{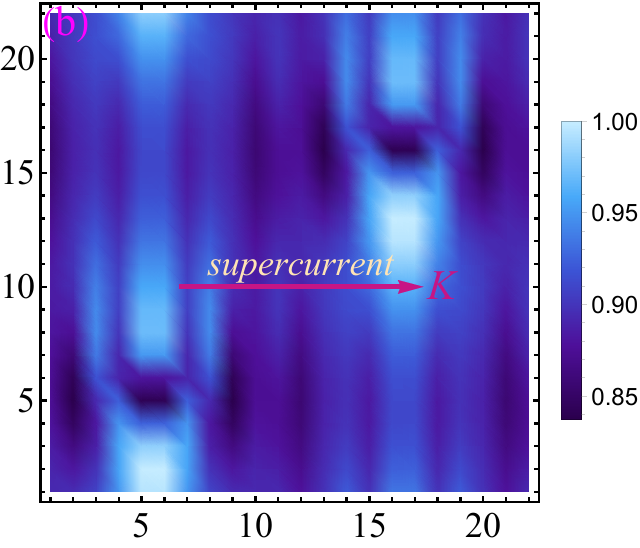}\label{fig:IntsK4}}
  \subfigure{\includegraphics[width=1.6in]{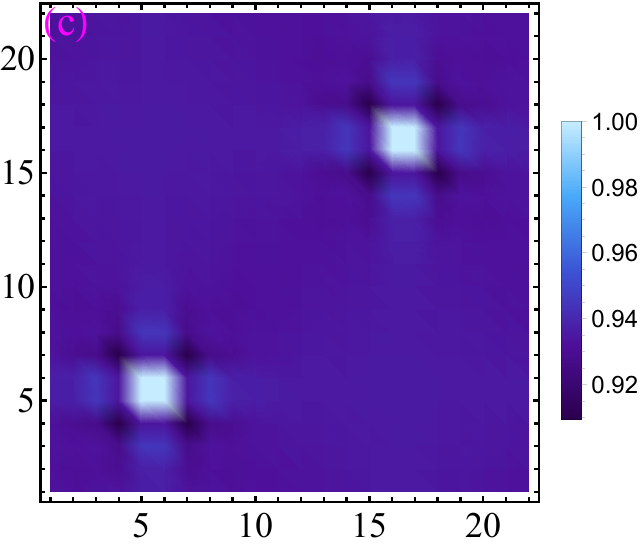}\label{fig:Intd}}~~
  \subfigure{\includegraphics[width=1.6in]{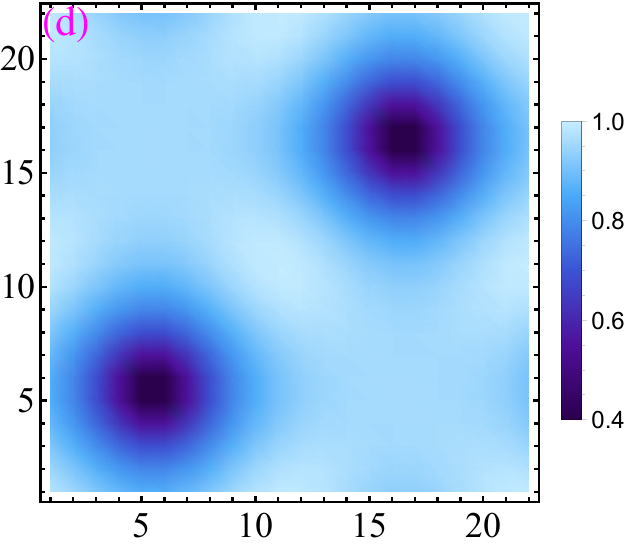}\label{fig:Intp}}
  \caption{The normalized intensity profile $|\Phi(x,y)|^2/|\Phi(x,y)|^2_{max}$ of zero-energy modes in a $22\times 22$ magnetic unit cell, with $\Delta_0=0.5$. Two vortices are located at ($5.5$, $5.5$) and ($16.5$, $16.5$), respectively. (a) The confined phase of the PB model. (b) The deconfined phase of the PB model, where the red arrow denotes the in-plane supercurrent in the $x$ direction, with $\mathit{K}=4\Delta_0$. (c) The $d_{x^2-y^2}$ pairing WeylSC. (d) The $p_x-ip_y$ pairing WeylSC. }\label{fig:Int}
\end{figure}
In this work, we also explore additional distinctions between MZMs and ZLLs. In Fig.~\ref{fig:Int}, we present the intensity profile of zero-energy modes on a magnetic unit cell for the confined and deconfined phases of the PB model, as well as for the heterostructure WeylSCs with $d_{x^2-y^2}$ and $p_x-ip_y$ pairing. Note that the ZLL exists only if $\mu=0$ and $k_z=\tfrac{\pi}{2}$. The intensity profile at the $(x,y)$ point is calculated by $|\Phi(x,y)|^2=\sum_{\bf k} |\psi_{\bf k}(x,y)|^2$, summed over the magnetic Brillouin zone. Here, we define a four-component spinor $\psi_{\bf k}(x,y) = \{\psi_{e,{\bf k},\uparrow}(x,y), \psi_{e,{\bf k},\downarrow}(x,y), \psi_{h,{\bf k},\uparrow}(x,y), \psi_{h,{\bf k},\downarrow}(x,y)\}^T$, where the subscripts $e$ and $h$ denote the electron and hole components, respectively. We can also construct an eigenvector $\Psi_{\bf k} = \{\psi_{\bf k}(x,y)\}^T$ for a zero-energy mode, which satisfies $\mathcal{H}_M({\bf k})\Psi_{\bf k}=0$, where $(x,y)$ traverses the magnetic unit cell and $\mathcal{H}_M({\bf k})$ is the momentum-space BdG Hamiltonian of WeylSCs in the presence of a vortex lattice. 

The normalized intensity profile of the MZM in the confined phase of the PB model is shown in Fig.~\ref{fig:IntsK0}. The local density of states (LDOS) is primarily distributed near the vortex cores, and the intensity nearly drops to zero away from the cores. In contrast, the intensity profile of the ZLL is delocalized, as shown in Figs.~\ref{fig:IntsK4},~\ref{fig:Intd} and~\ref{fig:Intp}. In the deconfined phase of the PB model, the supercurrent-induced oscillation of the LDOS is visible in our numerical results (Fig.~\ref{fig:IntsK4}), which show a pronounced periodic modulation in the $x$ direction. The intensity profile for ZLLs in heterostructure WeylSCs  is also delocalized but does not exhibit spatial oscillations. Remarkably, the intensity profile for the $d_{x^2-y^2}$ pairing state is nearly uniform (about $0.93$), except for slight enhancements near the vortex cores. In contrast, the LDOS shows significant declines when approaching the vortex cores for the $p_x-ip_y$ pairing, as shown in Fig.~\ref{fig:Intp}.

\begin{figure}[ht]
  \centering
  \subfigure{\includegraphics[width=3.2in]{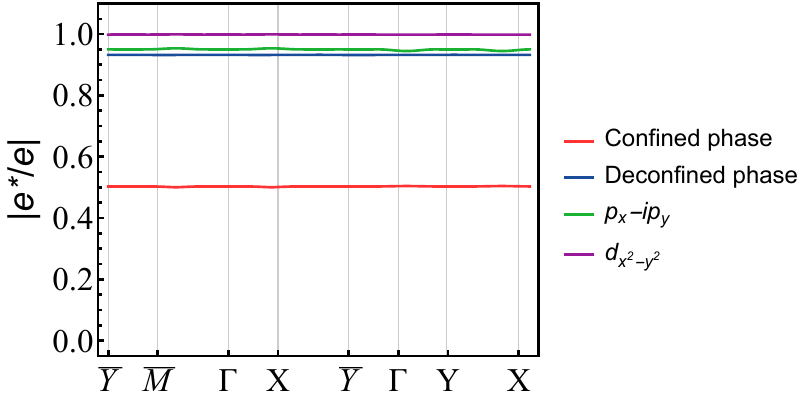}}
  \caption{The effective charge of the zero-energy modes in the vortex lattice of the PB model and WeylSCs, with $l_B=22$ and $\Delta_0=0.5$. Cooper pair momentum $\mathit{K}=4\Delta_0$ in the deconfined phase} \label{fig:Charge}
\end{figure}
Another distinction between MZMs and ZLLs is the effective charge. At a momentum point ${\bf k}$, the effective charge of a zero-energy mode is calculated by $C^e_{\bf k} =  \sum_{x,y,\varsigma} |\psi_{e,{\bf k},\varsigma}(x,y)|^2$, which indicates the proportion of electrons in Bogoliubov QPs, where $\varsigma = \uparrow, \downarrow$. Clearly, the proportion of holes is given by $C^h_{\bf k} =  \sum_{x,y,\varsigma} |\psi_{h,{\bf k}, \varsigma}(x,y)|^2 = 1-C^e_{\bf k}$. The numerically calculated effective charge along the momentum path, as plotted in Fig.~\ref{fig:BZ}, is shown in Fig.~\ref{fig:Charge}. We obtain $C^e_{\bf k} = 0.5$ for the MZM (red line), indicating that the MZM is charge-neutral due to $C^e_{\bf k} = C^h_{\bf k}$. In contrast, the calculated effective charge of the ZLLs is clearly far from $0.5$, revealing their non-charge-neutral nature. It should be noted especially that the ZLL is doubly degenerate, so our numerical calculations yield $C^e_{1,{\bf k}} = C^h_{2,{\bf k}}$, with $1$ and $2$ denoting the two degenerate ZLLs. The non-charge-neutrality of the ZLL has also been revealed in Ref.~\onlinecite{Pacholski.PRL2018} through an $s$-wave heterostructure WeylSC model. 
\section{Magneto-Optical conductivity}\label{sec:MOC}
In superconductors, $U(1)$ symmetry is broken, and charge is not conserved, which makes the detection of QP states, such as charge-neutral Majorana fermions, challenging from both experimental and theoretical points of view. However, an alternative approach to observe these states in superconductors is through optical response measurements. For instance, the local optical response has been proposed to detect dispersive chiral Majorana edge states in 2D topological superconductors~\cite{JamesPRL2021}. Theoretical calculations have shown that peaked magneto-optical conductivity, induced by Dirac-LLs of Bogoliubov QPs, can be obtained in the mixed state of a heterostructure WeylSC with $s$-wave pairing~\cite{ZhihaiLiuPRB2024}.

Here, we investigate the optical response of QPs in the mixed state for WeylSC models discussed in Sec.~\ref{sec:Model}. The longitudinal magneto-optical conductivity tensor $\sigma_{xx}$ in the vortex lattice can be derived from the Kubo formula~\cite{Ahn.NatureC2021,Tianrui.PRB2019,Kamatani.PRB2022,Papaj.PRB2022}. Expressed in the LL basis in the clean limit, we have
\begin{eqnarray}
\sigma_{xx}(\omega)=- \tfrac{ie^2}{2\pi l_B^2} \sum_{nn^{\prime}} \int \tfrac{d{\bf k}}{(2\pi)^3} \tfrac{f(E_n)-f(E_{n^\prime})}{E_n-E_{n^\prime}} \times \tfrac{V_{x}^{nn^\prime} V_{x}^{n^\prime n}}{\omega+E_n -E_{n^\prime} + i\delta},
\label{eqMOC}
\end{eqnarray}
where $f(\epsilon)=1/(1+e^{\epsilon/k_BT})$ is the Fermi distribution function and we set the Boltzmann constant $k_B \equiv 1$. We take $d{\bf k}=dk_xdk_ydk_z$ due to the possible dispersion of QP bands in the vortex lattice of WeylSCs. The velocity matrix element $V_{x}^{nn^\prime} = \langle \Psi_{n{\bf k}} |V_{x}| \Psi_{n^\prime{\bf k}} \rangle$ with $| \Psi_{n{\bf k}} \rangle$ being the eigenvector of the $n$-th LL (QP band), satisfying $\mathcal{H}_M|\Psi_{n{\bf k}} \rangle = E_n |\Psi_{n{\bf k}} \rangle$. The velocity operator in the superconducting state is given by~\cite{Ahn.NatureC2021}
\begin{eqnarray}
V_{x} ({\bf k}) = 
\begin{pmatrix}
\partial_{k_x} H^{\bf B}_0({\bf k}-e{\bf A}) & 0 \\
0 & \partial_{k_x} H_0^{{\bf B}*}(- {\bf k}-e{\bf A})
\end{pmatrix},
\end{eqnarray}
where $H^{\bf B}_0$ can be obtained from the BdG Hamiltonian $\mathcal{H}_M$.

\begin{figure}[ht]
  \centering
  \subfigure{\includegraphics[width=2.7in]{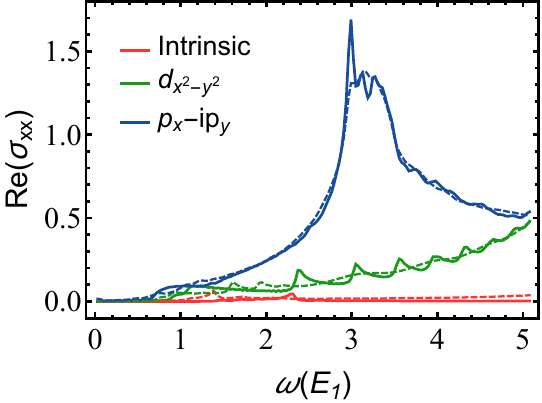}}
  \caption{The real part of the longitudinal magneto-optical conductivity (in units of $e^2/2\pi l_B^2$) of WeylSCs, with $l_B=10$, $\mu=0$, $T=0.001$ and the dimensionless $\delta=0.01$. $\Delta_0=1$ and $\Delta_0=0.5$ for intrinsic and heterostructure WeylSCs, respectively. The dashed lines indicate the magneto-optical conductivity of tilted Weyl cones with $\eta_x=0.5$; while the other parameters remain consistent.} \label{MOCond}
\end{figure}

The real part of the calculated magneto-optical conductivity is shown in Fig.~\ref{MOCond}, where ``intrinsic'' denotes the single-band WeylSC with $p_x-ip_y$ pairing. Here, we consider the tilt of the Weyl cones via a parameter $\eta_x$, which squeezes the space of LLs and breaks the usual dipolar selection rules $n\to n\pm 1$ for optical transitions~\cite{LukosePRL2007,GoerbigEPL2009,JuditPRB2015,TchoumakovPRL2016,YuZhi-MingPRL2016}. For singlet $d_{x^2-y^2}$ pairing, the Dirac-LLs give rise to a WSM-like magneto-optical conductivity, which exhibits a series of peaks at $\omega/E_1 = \sqrt{n}+\sqrt{n+1}$ ($n\ge 0$) because the major contribution to the conductivity comes from the $n\to n\pm 1$ transitions~\cite{VPGusynin.JPCM2007}, along with a linear background resulting from the dispersion of LLs in the $k_z$ direction~\cite{Ashby.PRB2013}. The tilt of Weyl cones shifts the conductivity peaks to lower optical frequencies. The magneto-optical conductivity for $d_{x^2-y^2}$ pairing is similar to that revealed in the $s$-wave heterostructure WeylSC~\cite{ZhihaiLiuPRB2024}. 

In contrast, the magneto-optical conductivity for the triplet $p_x-ip_y$ pairing shows only a hump, which is slightly suppressed when the BW cones are tilted, as shown in Fig.~\ref{MOCond} by the blue lines. A linear background exists at low frequencies, but there is no multiple-peak feature due to the absence of the flat Dirac-LLs $E_n=\sqrt{n}E_1$ for $n>0$. On the other hand, as displayed by the red lines, the magneto-optical conductivity of the intrinsic WeylSC does not exhibit a linear background because the MZM exists continuously when $K_-<k_z<K_+$ (see Fig.~\ref{fig:SpLLz}), and there is no apparent multiple-peak structure despite the gauge field ${\bf a}$ inducing pseudo-LLs $E_n=\sqrt{n} E_1/\alpha_D$ near the BW nodes. Meanwhile, a single peak emerges in the magneto-optical conductivity of the single-band WeylSC and shifts to a lower frequency when the tilt parameter $\eta_x \neq 0$.
\section{summary}\label{sec:summary}
In this work, we investigate the low-energy QP excitations and optical conductivity in the mixed state of WeylSCs realized through superconductor heterostructures with unconventional pairing symmetry, including the $d_{x^2-y^2}$ and $p_x-ip_y$ pairing. We reveal that QPs in the mixed state for the singlet $d$-wave pairing exhibit Dirac-LLs, similar to those demonstrated in the heterostructure WeylSC with $s$-wave pairing. In contrast, the triplet chiral $p$-wave pairing results in strongly dispersive QP bands, except for the chiral symmetry-protected, dispersionless ZLL. Consequently, there is no Dirac-LL structure in the excitation spectrum of the triplet $p_x-ip_y$ pairing state in the presence of a vortex lattice.

The QP excitations in the mixed state of heterostructure WeylSCs can be verified through the magneto-optical conductivity measurements. For spin-singlet $d_{x^2-y^2}$ pairing, the Dirac-LLs yield a WSM-like magneto-optical conductivity that displays a series of peaks superimposed on a linear background. In contrast, the conductivity curve for spin-triplet $p_x-ip_y$ pairing shows only a hump, with no evident conductivity peaks. Additionally,  we calculate the magneto-optical conductivity of the single-band WeylSC with $p_x-ip_y$ pairing, whose excitation spectrum in the vortex lattice simultaneously exhibits MZMs and pseudo-LLs. This calculation shows a single conductivity peak without a linear background.

We also investigate the distinctions between MZMs and ZLLs, both of which may be observed in the mixed state of heterostructure topological superconductors. We identify three key distinctions: (1) chiral symmetry is necessary for the dispersionless ZLLs but not for MZMs, in other words, when chiral symmetry is broken, ZLLs deviate from zero energy, while the topologically protected MZMs remain at zero energy; (2) MZMs are localized at the vortex core, whereas ZLLs are delocalized and extend over the magnetic unit cell; (3) MZMs are charge-neutral, unlike ZLLs, where the proportions of electrons and holes in Bogoliubov QPs are not equal. 
\acknowledgments
This work was supported by Guangdong Basic and Applied Basic Research Foundation (Grant
No. 2021B1515130007), the National Natural Science Foundation of China (Grant No. 12004442).

\normalem
\bibliography{QOC_in_WeylSCs}

\begin{thebibliography}{63}
\bibitem [{Ham()}]{HamiltSI}%
  \BibitemOpen
  \href@noop {} {}\bibinfo {note} {Hamiltonians~(\ref{HamSDBW}) and~(\ref{HamCPBW}) do not include virtual transitions to higher bands, of second order in $v_s$.}\BibitemShut {Stop}%
\end{thebibliography}

\end{document}